# *ZRT1* harbors an excess of nonsynonymous polymorphism and shows evidence of balancing selection in *Saccharomyces cerevisiae*

Elizabeth K. Engle* and Justin C. Fay§

*Molecular Genetics and Genomics Program, Washington University, St. Louis, MO, 63108

§Department of Genetics and Center for Genome Sciences and Systems Biology, Washington University, St. Louis, MO, 63108

Corresponding author:

Justin C. Fay

4444 Forest Park Blvd,

St. Louis, MO 63108

Phone: 314-747-1808

Fax: 314-362-2156

email: jfay@genetics.wustl.edu

## ABSTRACT

Estimates of the fraction of nucleotide substitutions driven by positive selection vary widely across different species. Accounting for different estimates of positive selection has been difficult, in part because selection on polymorphism within a species is known to obscure a signal of positive selection between species. While methods have been developed to control for the confounding effects of negative selection against deleterious polymorphism, the impact of balancing selection on estimates of positive selection has not been assessed. In *Saccharomyces cerevisiae*, there is no signal of positive selection within protein coding sequences as the ratio of nonsynonymous to synonymous polymorphism is higher than that of divergence. To investigate the impact of balancing selection on estimates of positive selection we examined five genes with high rates of nonsynonymous polymorphism in *S. cerevisiae* relative to divergence from *S. paradoxus*. One of the genes, a high affinity zinc transporter *ZRT1*, shows an elevated rate of synonymous polymorphism indicative of balancing selection. The high rate of synonymous polymorphism coincides with nonsynonymous divergence between three haplotype groups, which we find to be functionally indistinguishable. We conclude that balancing selection is not likely to be a common cause of genes harboring a large excess of nonsynonymous polymorphism in yeast.

# INTRODUCTION

The frequency of adaptive substitutions driven by positive selection is central to our understanding of molecular evolution and divergence between species. The neutral theory assume that the vast majority of substitutions are effectively neutral and generates predictions that can be tested based on patterns of molecular evolution (Fay and Wu 2003). While many individual genes have been found to deviate from neutral patterns of evolution, the overall impact of positive selection across the genome remains a contentious issue (Hahn 2008; Sella *et al*. 2009; Nei *et al*. 2010; Fay 2011).

Genome-wide comparisons of polymorphism versus divergence have been the primary means of estimating the frequency of positive selection between species. The McDonald-Kreitman (MK) test (McDonald and Kreitman 1991) has been used to estimate the frequency of positive selection within protein coding sequences based on an elevated ratio of nonsynonymous to synonymous divergence relative to that of polymorphism. However, applications of the MK test to plant, animal and microbial genomes have revealed substantial differences in estimates of positive selection among species, ranging from zero to over half of all amino acid substitutions (Fay 2011). While the frequency of positive selection may differ due to a species' effective population size and species-specific selective pressures (Bachtrog 2008; Gossmann *et al*. 2010; Siol *et al*. 2010; Slotte *et al*. 2010; Gossmann *et al*. 2012), estimating the frequency of positive selection during divergence between species depends on controlling for the effects of selection on polymorphism within species (Fay and Wu 2001; Bierne and Eyre-Walker 2004; Hughes *et al*. 2008).

Estimates of the frequency of positive selection can be influenced by a number of factors that can make it difficult to detect adaptation when it is present. Slightly deleterious polymorphisms segregate at low frequencies due to weak negative selection and can increase the ratio of nonsynonymous to synonymous polymorphism to a greater extent than that of divergence. As a consequence, deleterious polymorphism can obscure evidence of positive selection (Fay *et al*. 2002; Bierne and Eyre-Walker 2004; Charlesworth and Eyre-Walker 2008; Eyre-Walker and Keightley 2009). Methods have been developed to account for the effects of low frequency deleterious polymorphism; but even so, there are still some species with little to no evidence of positive selection (Fay 2011).

A number of other factors can influence the detection of positive selection through their effects on slightly deleterious polymorphism. Such factors include mating system as well as population size and structure. For example, a decrease in population size can increase the abundance of slightly deleterious polymorphism within a species and obscure evidence of positive selection between species (Eyre-Walker 2002). In humans there is little evidence for an excess of nonsynonymous divergence yet it has been estimated that up to 40% of amino acid substitutions could have been driven by positive selection without being detected (Eyre-Walker and Keightley 2009). Controlling for these additional factors is often difficult as it requires specific knowledge of the species being examined and its population history.

Another factor that has received less attention but can also influence estimates of positive selection is balancing selection (Wright and Andolfatto 2008). While all cases of balancing selection may not affect estimates of positive selection, those that maintain multiple nonsynonymous polymorphisms within a species could increase the genome-wide ratio of nonsynonymous to synonymous polymorphism within a species above that between species. Elevated rates of nonsynonymous polymorphism may also occur due to local adaptation (Charlesworth *et al*. 1997). If an appreciable number of genes are involved in adaptive divergence between different populations of the same species, genome-wide estimates of the frequency of positive selection between species could be substantially underestimated.

*Saccharomyces cerevisiae* is one species with little to no evidence of positive selection based on the MK test (Doniger *et al*. 2008; Liti *et al*. 2009; Elyashiv *et al*. 2010). In contrast to other species that lack evidence of positive selection (Foxe *et al*. 2008; Gossmann *et al*. 2010; Gossmann *et al*. 2012), its large effective population size ensures the efficient removal of weakly deleterious mutations and the ability to fix weakly advantageous mutations. However, *S. cerevisiae* also exhibits strong population structure, potentially facilitated by its low rate of outcrossing (Ruderfer *et al*. 2006), low rate of migration, or local adaptation to the diverse array of environments from which it has been isolated (Fay and Benavides 2005). Genome-wide patterns of population structure have revealed a number of genetically differentiated groups, including strains originating from sake in Japan, vineyards in Europe and oak trees in North America (Liti *et al*. 2009; Schacherer *et al*. 2009). While these groups may have arisen due to geographic barriers, they might also have arisen as a consequence of domestication or adaptation to human modified environments (Fay and Benavides 2005). However, even when these groups are taken into consideration and examined separately, the ratios of nonsynonymous to synonymous polymorphism within or between groups are higher than the ratio of nonsynonymous to synonymous divergence between species (Elyashiv *et al*. 2010).

In this study, we tested the hypothesis that genes with a large excess of nonsynonymous polymorphism are under balancing selection. We reasoned that such genes have a disproportionate effect on estimates of positive selection and should be considered separately if under balancing selection. We examined five genes that were previously shown to contain a large excess of nonsynonymous to synonymous polymorphism (Doniger *et al*. 2008; Liti *et al*. 2009). To distinguish between purifying selection and balancing selection on nonsynonymous polymorphism we examined rates of synonymous polymorphism since negative selection is expected to decrease linked neutral variation whereas balancing selection is expected to increase linked neutral variation (Charlesworth *et al*. 1997). We found one of the genes, *ZRT1*, shows a significantly elevated rate of synonymous polymorphism based on the Hudson-Kreitman-Aguade (HKA) test (Hudson *et al*. 1987), consistent with balancing selection. Our results show that a large number of amino acid polymorphisms can occur at certain loci under balancing selection, but that such loci are not particularly common in yeast.

## MATERIALS AND METHODS

**Polymorphism and divergence data:** Data were collected for five genes that were previously found to exhibit an excess of nonsynonymous polymorphism in two studies (Doniger *et al*. 2008; Liti *et al*. 2009) and 30 randomly selected control genes using 36 *S. cerevisiae* strains with genome sequence data (Table S1). Twenty-seven of the genome sequences were accessed through a blast server (www.moseslab.csb.utoronto.ca/sgrp/) and the other nine were accessed through the Saccharomyces Genome Database (www.yeastgenome.org). For each gene, sequences homologous to the coding region of the reference genome (S288C) were aligned using ClustalX (2.0) (Larkin *et al*. 2007). Strains with sequences that were <99% and <90% of the S288C sequence length for the neutral and selected genes, respectively, were removed. Strains were removed from a gene analysis if a polymorphism led to an internal stop codon (4 cases) while unique single base insertions were considered sequencing errors and the base was removed from the sequence (10 cases). A small number of heterozygous sites were present within Vin13, VL3 and LalvinAQ23. At these sites, we randomly selected one of the two observed nucleotides to represent the position. Divergence was measured by comparison to the CBS432 strain of *S. paradoxus*.

The final dataset included an average of 29.9 strain alleles per gene, ranging from 23 to 36, and the five gene set included an average of 21.4 strain alleles per gene, ranging from 18 to 24. Eight of the control genes were removed from the analysis because they had few strains with sufficient sequence coverage or multiple strains with frameshifts. One gene, *RPS28B*, was removed since it showed evidence of introgression between species (Doniger *et al*. 2008).

**Population genetic analysis:** MK tests were conducted using the number of nonsynonymous and synonymous polymorphic sites and fixed differences calculated using DnaSP (v.5.10.01) (Librado and Rozas 2009). The weighted neutrality index (Stoletzki and Eyre-Walker 2011) was estimated by:

$$\mathrm{NI} = \frac{\sum \frac{D_{si} P_{ni}}{(P_{si} + D_{si})}}{\sum \frac{P_{si} D_{ni}}{(P_{si} + D_{si})}}$$

where *P* and *D* indicate the number of polymorphic sites and fixed differences, respectively, subscripts *s* and *n* indicate synonymous and nonsynonymous changes, respectively, and *i* indicates the *i*th gene.

HKA tests were conducted using MLHKA v2.0 (Wright and Charlesworth 2004) using rates of synonymous polymorphism and divergence obtained from DnaSP (Table S2). Each of the five genes significant by the MK test were compared to 21 control genes, covering 22,974 sites, using the MLHKA test. The program was run with a chain length of 100,000 for all analyses.

For analysis of the region surrounding *ZRT1,* from *MNT2 t*hrough *FZF1* (~11kb), we downloaded *S. cerevisiae* and *S. paradoxus* strain sequences from the Saccharomyces Genome Resequencing Project (Liti *et al*. 2009). A sliding window analysis of polymorphism and divergence was calculated in DnaSP using 36 *S. cerevisiae* strains and one *S. paradoxus* strain, CBS432, with gaps in the alignment excluded. Due to difficulty in aligning the 4.67kb noncoding region between ADH4 and *ZRT1*, we only used ~200 bases downstream of *ADH4* and 800 bases upstream of *ZRT1* where we were confident of the alignment.

A bootstrapped neighbor-joining tree for *ZRT1* and the concatenated control gene set were constructed using MEGA5 and pairwise gap-removal (Tamura *et al*. 2011).

**Strain construction and phenotype analysis:** *ZRT1* was deleted in YJF186 (YPS163 background, *Mat a*, HO*::dsdAMX4*, *ura3-140*) using the kanMX deletion cassette (Wach *et al*. 1994). Three *ZRT1* alleles were integrated into this strain at the *ura3* locus by amplifying the entire *ZRT1* gene region, including 878 bases of the 5' noncoding region and the entire 195 bases of the 3' noncoding region as well as 186 bases of the 3' gene *FZF1*, using primers with homology to pRS306 and transforming the product along with the yeast integrative plasmid, pRS306 (Sikorski and Hieter 1989). Integration of these constructs at the *ura3* locus was achieved by selection on plates lacking uracil and each transformant was confirmed by PCR. The *ZRT1* alleles were found to have between 1 and 3 mutations. However, most had only single synonymous changes or changes within the 5' or 3' regions and no mutations were shared among the alleles, including the replicated transformants. These mutations were considered not functional due to the lack of any phenotypic effects. Wild-type (YJF186) and *ZRT1* deletion strains were integrated with the empty plasmid pRS306 as a control.

Experiments comparing growth in low zinc conditions were conducted using low zinc media (LZM) composed of 0.17% yeast nitrogen base without amino acids, $(NH)_2SO$, or zinc (MP Biomedicals); 0.5% $(NH4)_2SO_4$; 20 mM trisodium citrate, pH 4.2; 2% glucose; 1 mM $Na_2EDTA$; 25 uM $MnCl_2$; and 10 uM $FeCl_3$, as previously described (Gitan *et al*. 1998; Gitan *et al*. 2003). Strains were grown overnight in LZM, washed, diluted to a starting OD of 0.05 in fresh LZM with 0.1 mM of $ZnCl_2$, and then grown for 20 hours in an iEMS plate reader at 30° with 1200 rpm shaking (model no. 1400; Thermo Lab Systems, Helsinki, Finland). For each strain, the maximum OD was determined after normalization to the initial cell concentration. For each *ZRT1* construct and controls, 3 to 9 independent transformants were phenotyped.

Rates of fermentation were measured using grape juice. Strains were grown overnight in Reserve Chardonnay grape juice (Winexpert, Port Coquitiam, B.C., Canada), washed and diluted to a starting OD of 0.1 in fresh grape juice or grape juice with metal chelators (20 mM trisodium citrate, pH 4.2 and 1 mM $Na_2$EDTA). The fermentation was conducted in 250 mL flasks, sealed with airlocks and incubated at room temperature out of direct sunlight without shaking. Flasks were weighed daily to determine $CO_2$ loss and shaken once daily, immediately following measurement. Four independent transformants were examined for each construct.

## RESULTS

**Identification of genes exhibiting an excess of amino acid polymorphism:** From two previous independent genome-wide screens based on the McDonald-Kreitman (MK) test, we identified five genes that were significant (P < 0.001) in both studies (Doniger *et al*. 2008; Liti *et al*. 2009). The five genes are *IRA2*, *OPT2*, *PEP1*, *SAS10* and *ZRT1*. All of the genes show a ratio of nonsynonymous to synonymous polymorphism (Pn/Ps) that is more than two-fold greater than that of divergence (Dn/Ds). We repeated the MK test using a strain set comprised of 36 strains for which genome sequences are publicly available (see Methods and Table S1), of which 29 were not included in the previous screens. All five genes retained significance by the MK test (P < 0.05, Bonferroni corrected, Table 1). As a control, we randomly selected 21 genes from those not significant in either previous study. Only two of the genes were significant by the MK test (P < 0.05, Bonferroni corrected, Table S2), both characterized by Pn/Ps greater than Dn/Ds.

The two gene sets exhibited marked differences in their neutrality index (Table 1). The weighted neutrality index (Stoletzki and Eyre-Walker 2011) is 3.80 for the five selected genes and 1.33 for the 21 control genes. The high neutrality index of the five genes, indicating an excess of nonsynonymous polymorphism, is unlikely a consequence of selective pressure on synonymous sites since average codon bias is similar between the two groups and the five genes have nearly equal numbers of changes to preferred and unpreferred codons for both polymorphism, 63 and 58 respectively, and divergence, 414 and 397 respectively. Thus, the five gene set is highly enriched for genes with an excess of nonsynonymous polymorphism relative to divergence.

**Balancing selection in *ZRT1*:** A high ratio of nonsynonymous to synonymous polymorphism can result from slightly deleterious nonsynonymous mutations that contribute to polymorphism but not divergence or from a recent loss of functional constraint. In either scenario, the rate of synonymous polymorphism should not be affected. Alternatively, a high rate of nonsynonymous polymorphism can result from balancing selection on multiple nonsynonymous alleles. If balancing selection is responsible for the elevated rate of nonsynonymous polymorphism, rates of linked synonymous polymorphism should also be elevated (Charlesworth *et al*. 1997).

To test whether the rate of synonymous polymorphism is elevated in any of the five genes we used the Hudson-Kreitman-Aguade (HKA) test (Hudson *et al*. 1987). Using a maximum likelihood, multilocus HKA test (MLHKA), only *ZRT1* showed a significantly elevated rate of synonymous polymorphism in comparison to the control gene set (Table S3). Figure 1 shows that of all the genes we tested, *ZRT1* is characterized by an exceptionally high rate of synonymous polymorphism. One gene in the control gene set, *TRX2*, appears to be an outlier characterized by both a high rate of synonymous polymorphism as well as a low rate of synonymous divergence. *TRX2* is nominally significant for an excess synonymous polymorphism relative to the remaining neutral gene set by the MLHKA test (P = 0.0132). However, removal of *TRX2* from the neutral gene set only increases the significance of *ZRT1*. Thus, of the five genes, only *ZRT1* exhibits evidence of balancing selection.

Balancing selection on *ZRT1* is also supported by patterns of nonsynonymous and synonymous divergence among strains. A neighbor-joining tree of *ZRT1* shows that all of the *ZRT1* alleles, except for the EC1118 allele, cluster into three groups distinguished by multiple nonsynonymous and synonymous differences (Figure 2). With the exception of the wine strain group, the groups show no close correspondence to the source from which each strain was obtained or to a neighbor-joining tree generated from the concatenated control gene set (Figure 2 and Figure S1). Of the 111 polymorphic sites used to generate the tree, 86 can be placed on a single branch without homoplasies, of which 24 nonsynonymous and 17 synonymous changes occur on one of the four main internal branches (Figure 2). In comparison, 62 synonymous but only 12 nonsynonymous differences separate *S. cerevisiae* from *S. paradoxus*. Thus, the high ratio of nonsynonymous to synonymous polymorphism is not limited to external branches, as would be expected to occur if most nonsynonymous polymorphisms were deleterious.

The presence of intermediate frequency alleles, many of which contribute to the unique grouping of *ZRT1* alleles, also supports balancing selection. For synonymous sites Tajima's D is positive across all strains ($D = 0.718$, $P > 0.10$), but negative within each of the three *ZRT1* strain groups (M22 Group $D = -1.265$, YPS163 Group $D = -0.59894$, S288C Group $D = -0.61182$, all $P > 0.10$). In comparison, the average D of the control gene set is -0.294 and only 4 of the genes have positive D values greater than 0.3. These results further highlight the unique pattern of variation present at *ZRT1*.

**Regional variation around *ZRT1*:** The elevated rate of synonymous polymorphism in *ZRT1* could be a consequence of balancing selection on *ZRT1* amino acid polymorphism, but could also be caused by selection on its promoter or on adjacent genes. To determine whether the signal of balancing selection extends into adjacent genes and gene regions we applied the MLHKA test to the two genes adjacent to *ZRT1*, *ADH4* and *FZF1*. Only *ADH4* was significant in comparison to the control gene set (MLHKA test, $P = 0.0007$, Table S3) and was characterized by both high rates of polymorphism but also low rates of divergence at synonymous sites (Figure 1). Hence, we also tested the next gene adjacent to *ADH4*, *MNT2*, and found no significant departure from neutrality as measured by the MLHKA test. Additionally, none of the three adjacent genes that were examined showed a significant excess of amino acid polymorphism as measured by the MK test (Table S2).

To more precisely track the signal of balancing selection within and around *ZRT1*, we used a sliding window analysis of polymorphism to divergence, including both coding and noncoding regions. Figure 3 shows that the highest rate of polymorphism occurs within the coding region of *ZRT1* and extends into its 5' noncoding region. The overall rate of polymorphism is much lower in the two adjacent genes *ADH4* and *FZF1*. In *ADH4*, the rate of divergence is also quite low and likely contributes to the significance of the MLHKA test. Interestingly, a portion of *MNT2* has a very low rate of polymorphism whereas its more distal portion has another peak of polymorphism. Based on the sliding window analysis and the MLHKA test results, the signature of balancing selection appears to be concentrated at the *ZRT1* locus.

We next examined the degree to which polymorphism within *ZRT1* is independent of polymorphism within adjacent genes. There is ample evidence of recombination within and around *ZRT1*. Across the entire region, from *MNT2* through *FZF1*, there have been a minimum of 26 recombination events based on the four-gamete test (Hudson and Kaplan 1985). As expected in the presence of recombination, the genealogies of *ADH4* and *FZF1* differ from that of *ZRT1* (Figure S2), though all three genes show a similar grouping of wine strains. *ADH4* is the most similar to *ZRT1* but has less divergence. As measured by the HKA test, *ZRT1* shows significantly elevated rates of polymorphism compared to *FZF1* ($P = 0.0087$), but not *ADH4* or *MNT2* ($P > 0.05$).

***ZRT1* alleles confer no detectable phenotype differences:** If selection has acted on *ZRT1* then different alleles of *ZRT1* should confer different phenotypes. *ZRT1* is a high affinity zinc transporter that is activated only when zinc levels are very low and facilitates growth under limiting zinc conditions (Zhao and Eide 1996). We compared the effects of three *ZRT1* alleles integrated into a strain in which the endogenous *ZRT1* gene was deleted. The three alleles were from the S288C (laboratory), M22 (wine), and YPS163 (nature) strains and were selected as representatives from the three major groups of strains (Figure 2). While deletion of *ZRT1* resulted in a significant growth defect in zinc limiting conditions and each of the three alleles rescued the growth deficiency, we found no significant difference among the three *ZRT1* alleles for maximum growth (Figure 4) or growth rate (not shown).

In addition to its requirements for growth, zinc is an essential cofactor for many enzymes, including alcohol dehydrogenase, and has been shown to influence rates of fermentation (De Nicola and Walker 2011). To test whether *ZRT1* alleles affect rates of fermentation we measured $CO_2$ release during fermentation of grape juice into wine. In the presence of metal chelators, deletion of *ZRT1* had a dramatic effect on the rate of fermentation; but no differences were found among the three *ZRT1* alleles tested (Figure 5). No differences in rates of fermentation were found among any of the four strains in grape juice without chelators.

The ability of each *ZRT1* allele to rescue the *ZRT1* deletion phenotypes indicates that none of the 38 amino acid polymorphisms that distinguish these three *ZRT1* alleles cause a substantial loss of function as measured by growth or fermentation rate under the conditions assayed.

## DISCUSSION

Application of the MK test to a variety of species has revealed substantial differences in the estimated frequency of positive selection on protein coding sequences (Fay 2011). While differences in effective population size are capable of explaining some of the differences among species (Eyre-Walker and Keightley 2009; Gossmann *et al*. 2010; Halligan *et al*. 2010; Siol *et al*. 2010; Slotte *et al*. 2010; Gossmann *et al*. 2012), a small effective population cannot explain the absence of evidence for positive selection in yeast. In this study, we examined whether balancing selection can explain the high rate of nonsynonymous polymorphism observed in a small set of genes exhibiting a disproportionately large excess of nonsynonymous polymorphism, as this could obscure evidence of positive selection in *S. cerevisiae*. We showed that one out of the five genes tested exhibits a significantly elevated rate of synonymous polymorphism indicative of balancing selection. While patterns of polymorphism and divergence around *ZRT1* suggest that nonsynonymous polymorphism within *ZRT1* itself is the most likely target of balancing selection, we found no functional differences among three alleles using two different phenotype assays. Our results provide a case of balancing selection in yeast and indicate that balancing selection is not likely to be common, at least among those genes with disproportionately large effects on estimates of positive selection.

**Balancing selection at *ZRT1*:** Evidence for balancing selection on *ZRT1* is based on an elevated rate of synonymous polymorphism as measured by the HKA test (Figure 1 and Table S3), a high ratio of polymorphism to divergence that is centered on *ZRT1* (Figure 3), an increased frequency of intermediate frequency alleles and the coincidence of multiple synonymous and nonsynonymous changes that distinguish three groups of strains (Figure 2). However, it is worth noting that balancing selection in the general sense, i.e. selective maintenance of distinct alleles, can result from temporal or spatial variation in selection coefficients as well as heterozygote advantage. Local adaptation, which can result from spatial variation in selection coefficients, also provides an explanation for the presence of multiple nonsynonymous differences among alleles. Yet, our results are not able to distinguish between these different forms of selection, but rather distinguish them from patterns that can be explained by population structure, loss of selective constraint, and selection on adjacent genes.

In *S. cerevisiae*, there is extensive population structure, related to both geographic origin and the ecological source from which each strain was isolated (Fay and Benavides 2005; Liti *et* *al*. 2009; Schacherer *et al*. 2009), which are frequently correlated with one another. Such groups include the sake strains from Japan, oak tree strains from North America and strains isolated from Europe or vineyards. The neighbor-joining tree of the 21 control genes generally recapitulates these previously defined groups. While the *ZRT1* tree bears some resemblance to that of the control gene set, particularly the vineyard group, the three main groups of strains differentiated at *ZRT1* are not obviously related by either their geographic origin or ecological source from which they were isolated. More importantly, population structure by itself does not explain the elevated rates of polymorphism at *ZRT1* relative to the 21 control gene set. In support of selection acting at *ZRT1*, we observed negative Tajima's D values for the majority of the control gene set but a positive Tajima's D value at *ZRT1*, consistent with balancing selection.

Loss of functional constraint or weak negative selection is another explanation for an excess of nonsynonymous polymorphism as measured by the MK test. Genome-wide estimates in yeast suggest that much of the nonsynonymous polymorphism may be weakly deleterious (Elyashiv *et al*. 2010). In the case of *ZRT1*, we cannot exclude the possibility that some of the nonsynonymous polymorphisms are neutral or slightly deleterious. However, two lines of evidence indicate that at least some of the nonsynonymous changes within *ZRT1* have been under balancing selection. First, many neutral and most deleterious polymorphisms are expected to be rare and only present in a small number of strains. While 78% of nonsynonymous alleles are at less than 10% frequency in the 21 control gene set, only 38% of nonsynonymous polymorphism are at less than 10% frequency in *ZRT1*. In addition, of the nonsynonymous changes that are specific to one or more lineages, 55% are positioned along the four internal branches that distinguish the three major groups of *ZRT1* alleles. Second, neither loss of constraint or negative selection on nonsynonymous polymorphism should increase variation at linked synonymous sites.

While patterns of variation within and around *ZRT1* indicate that it is the most likely target of balancing selection, selection on linked sites could have influenced observed patterns of variation at *ZRT1*. Patterns of polymorphism within the adjacent gene, *FZF1*, indicate no excess of synonymous or nonsynonymous polymorphism. However, *FZF1* may have experienced a recent selective sweep since there is evidence of positive selection during *S. cerevisiae* and *S. paradoxus* divergence, both within its coding and within the intergenic region between *ZRT1* and *FZF1* (Sawyer *et al*. 2005; Engle and Fay 2012). Patterns of polymorphism within the adjacent genes *ADH4* and *MNT2* are more complex. Both genes show rates of synonymous polymorphism that are higher than the 21 control gene set, except for the outlier gene *TRX2*. *MNT2* shows regions with high and low polymorphism levels, but the region closest to *ZRT1* has the lower rate of polymorphism (Figure 3). *ADH4* shows a significantly elevated rate of synonymous polymorphism relative to divergence by the HKA test. Yet, in comparison to other regions (Figure 3) the significance of *ADH4* appears to be a partial consequence of the low rate of synonymous divergence. These observations combined with the ample evidence for recombination within the region indicate that while sites within *MNT2* and *ADH4* may have also been under selection, selection on linked sites in adjacent regions are unlikely to be solely responsible for the high rate of synonymous polymorphism at *ZRT1*.

Relevant to the possibility of selection on adjacent genes, there are functional links between *ADH4* and *ZRT1*. *ADH4* and *ZRT1* are both activated by *ZAP1* in zinc limiting conditions (Lyons *et al*. 2000), and *ADH4* is an alcohol dehydrogenase that may help conserve zinc or work more efficiently under zinc limiting conditions (Bird *et al*. 2006; De Nicola *et al*. 2007), or during fermentation of sugars to ethanol (Zhao and Bai 2012). Interestingly, the closest homologs of *ZRT1* outside of those present within the *sensu strictu Saccharomyces* species are from two distantly related species commonly found in wine fermentations, *Lachancea thermotolerans* and *Zygosaccharomyces rouxii* (Combina *et al*. 2005; Romancino *et al*. 2008), rather than other more closely related species, suggesting that *ZRT1* may have been introgressed into the ancestral lineage of the *Saccharomyces* species. The subtelomeric physical location of *ZRT1* in *S. cerevisiae* is consistent with other genes acquired by horizontal gene transfer (Hall *et al*. 2005; Muller and McCusker 2011) and genes likely to be involved with adaptations to specific environments (Brown *et al*. 2010).

**Phenotypic effects of *ZRT1* alleles:** We found three distinct *ZRT1* alleles all rescued two *ZRT1* deletion phenotypes but were not different from one another. This result indicates that, in the conditions tested, none of the nonsynonymous differences among the three alleles cause a substantial loss of *ZRT1* function. The lack of phenotypic differences among the different *ZRT1* alleles implies that either the alleles are functionally equivalent to one another, and so are not involved in balancing selection, or that the lack of a discernible phenotype is a consequence of the conditions tested or an effect too small to be detected. *ZRT1*, as a metal transporter, could influence fitness due to transport of other metals, such as cadmium (Gitan *et al*. 1998; Gomes *et al*. 2002; Gitan *et al*. 2003).

**The prevalence of balancing selection:** Out of the five genes that exhibit an excess of nonsynonymous polymorphism by the MK test, only *ZRT1* shows evidence of balancing selection. The excess of nonsynonymous polymorphism in the other four genes is most likely a consequence of loss of functional constraint or slightly deleterious polymorphism. Interestingly, alleles of *IRA2*, a GTPase that negatively regulates RAS signaling, are responsible for numerous environment-specific differences in gene expression across the genome (Smith and Kruglyak 2008). However, *IRA2* does not show an excess of synonymous polymorphism as measured by the HKA test.

The prevalence of balancing selection across the entire yeast genome is more difficult to assess. The observation that rates of nonsynonymous and synonymous polymorphism are correlated with one another provides some evidence for the possibility of weak balancing selection throughout the yeast genome (Cutter and Moses 2011). However, genes with high rates of synonymous polymorphism do not show a tendency towards an excess of nonsynonymous polymorphism (Kendall's tau = -0.15, Figure 6) as predicted by the MK test using the data of Liti *et al.* (2009). The challenge to interpreting genome-wide evidence for balancing selection is that many cases of balancing selection may be difficult to detect. First, the effect of balancing selection on linked variation decreases as a function of the rate of recombination; nucleotide diversity is $1+1/4Nr(1-F)$ relative to a neutral locus, where $N$ is the effective population size and $r$ is the rate of recombination and $F$ is the inbreeding coefficient (Charlesworth *et al*. 1997). Using a rate of recombination of $3.5 \times 10^{-6}$/bp, a rate of outcrossing of $2 \times 10^{-5}$/generation, and an effective population size of $1.6 \times 10^{7}$ (Ruderfer *et al*. 2006), we expect diversity to be increased by a factor of ten and two, 13 bp and 113 bp from a site under balancing selection, respectively. Gene conversion is expected to narrow this window even further (Andolfatto and Nordborg 1998). Second, balancing selection must act over many generations, on the order of the effective population size (Navarro *et al*. 2000), in order to noticeably influence linked neutral variation. However, the ability to detect balancing selection may be increased if there are multiple selected sites at a single locus, which might be the case for genes identified by a high rate of nonsynonymous polymorphism. Thus, it is hard to rule out the possibility that balancing selection has inflated the rate of nonsynonymous polymorphism across many genes without generating a strong effect on linked synonymous sites.

**Why is there little evidence of adaptive evolution within the yeast genome?** An important and persistent question in genome-wide estimates of adaptive evolution based on the MK test is why some species show high rates of adaptive evolution whereas others, such as yeast, do not. A small effective population size is one explanation since adaptive substitutions are expected to be more infrequent and deleterious polymorphism more common. A small effective population provides a reasonable explanation for the absence of signal in humans and many plant species (Eyre-Walker and Keightley 2009; Gossmann *et al*. 2010; Halligan *et al*.

2010; Siol *et al*. 2010; Slotte *et al*. 2010; Gossmann *et al*. 2012). However, it does not explain the lack of signal in yeast which has a large effective population size, on the order of $10^7$ for *S. paradoxus* (Tsai *et al*. 2008) and *S. cerevisiae* (Ruderfer *et al*. 2006). The rate of outcrossing may also be relevant to detecting selection in yeast. Selfing helps purge recessive deleterious alleles but also limits recombination between different haplotypes. Despite the presence of selfing in yeast, *S. cerevisiae* exhibits an excess of rare nonsynonymous polymorphism indicative of deleterious alleles and a rapid decay in levels of linkage disequilibrium, an observation that can be attributed to its exceptionally high rate of recombination even if outcrossing is rare. Thus, there is no obvious aspect of *S. cerevisiae* diversity that distinguishes it from outcrossing species. Further, both a selfing and outcrossing species of *Arabidopsis* show no signal of adaptive evolution (Foxe *et al*. 2008). As it stands, neither population size nor selfing provide a particularly compelling explanation for why yeast show little adaptive evolution based on the MK test.

In the present study, we considered the possibility of balancing selection obscuring patterns of positive selection in yeast. While we only focused on a small number of genes exhibiting a large excess of nonsynonymous polymorphism, we found one that exhibited evidence of balancing selection. We conclude that balancing selection is either too rare to be dominant factor influence evidence of adaptive evolution in yeast, or it does and is just not detectable. While not emphasized here, it is also important to consider whether adaptive evolution is rare but estimates of positive selection are inflated in some species (Fay 2011).

## SUPPORTING INFORMATION

Supporting information includes supplementary Tables S1-4 and Figures S1-2.

## ACKNOWLEDGEMENTS

This work was supported by a grant from the National Institutes of Health (GM080669).

## TABLES

Table 1. McDonald-Kreitman (MK) test results.

| Gene(s) | Number of sites | Pn[a] | Ps[a] | Dn[b] | Ds[b] | MK P-value | Neutrality index |
|---|---|---|---|---|---|---|---|
| *IRA2* | 9,117 | 69 | 108 | 131 | 735 | 0.000000 | 3.6 |
| *OPT2* | 2,436 | 35 | 20 | 18 | 187 | 0.000000 | 18.2 |
| *PEP1* | 4,293 | 48 | 45 | 168 | 320 | 0.002240 | 2.0 |
| *SAS10* | 1,680 | 12 | 7 | 41 | 120 | 0.002186 | 5.0 |
| *ZRT1* | 1,113 | 55 | 45 | 12 | 62 | 0.000000 | 6.3 |
| 5 genes[c] | 18,639 | 219 | 225 | 370 | 1424 | | 3.8 |
| 21 genes[c] | 22,815 | 158 | 260 | 711 | 1623 | | 1.3 |

[a]Pn and Ps are the number of nonsynonymous (Pn) and synonymous (Ps) polymorphic sites.

[b]Dn and Ds are the number of nonsynonymous (Dn) and synonymous (Ds) fixed differences.

[c]The five genes and 21 control genes are described in the text.

## FIGURE LEGENDS

**Figure 1. Synonymous polymorphism versus divergence.** Synonymous nucleotide diversity ($\pi_{SYN}$) versus synonymous nucleotide divergence ($K_{SYN}$) is shown for the five selected genes (red), the 21 control genes (black), and for three genes neighboring *ZRT1* (blue).

**Figure 2. Neighbor-joining tree of *ZRT1*.** A neighbor-joining tree of *ZRT1* is shown along with bootstrap values greater than 90% (gray). *S. cerevisiae* strains are color coded by class (see legend). The position of the branch leading to *S. paradoxus* is indicated by a dashed line, not drawn to scale. The number of nonsynonymous/synonymous/complex (two changes within a codon) changes unique to each of the four main lineages are listed along the respective branches. The inset on the right shows the unrooted neighbor-joining tree of the concatenated 21 control gene set drawn to the same scale and for the same strains as the *ZRT1* tree.

**Figure 3. Sliding window analysis of polymorphism and divergence within and around *ZRT1*.** The sliding window plot includes *ZRT1* and three neighboring genes with their positions and orientations indicated below the graph. Polymorphism (solid line) and divergence from *S. paradoxus* (dashed line) are shown for a window size of 200 bp and step size of 50 bp. A break is shown between *ADH4* and *ZRT1* where ~3,600 intergenic bases were excluded due to uncertainty in the alignment with *S. paradoxus*. The average nucleotide diversity of synonymous sites for the control gene set is indicated by the gray, horizontal line.

**Figure 4. Effects of strain-specific *ZRT1* alleles on growth in low-zinc conditions.** The maximum cell density in low-zinc conditions is shown for YPS163 with an unmodified *ZRT1* allele (WT), the same strain with a deletion of *ZRT1* (deletion), and three *ZRT1* alleles

integrated into the *ZRT1* deletion strain (YPS163, S288C and M22). The three integrated alleles represent alleles from the three major strain groupings based on *ZRT1*. Error bars show the 95% confidence interval of the mean.

**Figure 5. Effects of strain-specific *ZRT1* alleles on fermentation rate.** Fermentation rate, measured by $CO_2$ release in grams per hour, for strains grown in grape juice containing metal chelators. All strains have *ZRT1* deleted and three have either a S288C (orange), YPS163 (yellow) or M22 (green) allele of *ZRT1* inserted at the *URA3* locus. Lines show the average of four replicates. Standard deviations are not shown for clarity and average between 0.0028 and 0.0056 for the four strains.

**Figure 6. No abundance of genes exhibiting high rates of synonymous polymorphism and an excess of nonsynonymous polymorphism.** The rate of synonymous polymorphism, measured by the number of synonymous SNPs per codon, compared to the observed minus the expected number of nonsynonymous SNPs for 4992 genes from (Liti *et al*. 2009). The expected number of nonsynonymous SNPs is Pn - Ps*Dn/Ds, where Pn and Ps are the number of nonsynonymous and synonymous SNPs, respectively, and Dn and Ds are the number of nonsynonymous and synonymous fixed differences, respectively. Genes with nonsynonymous polymorphism below -10 or above 10 are shown as points at those values. The red point is *ZRT1*.

Figure 1

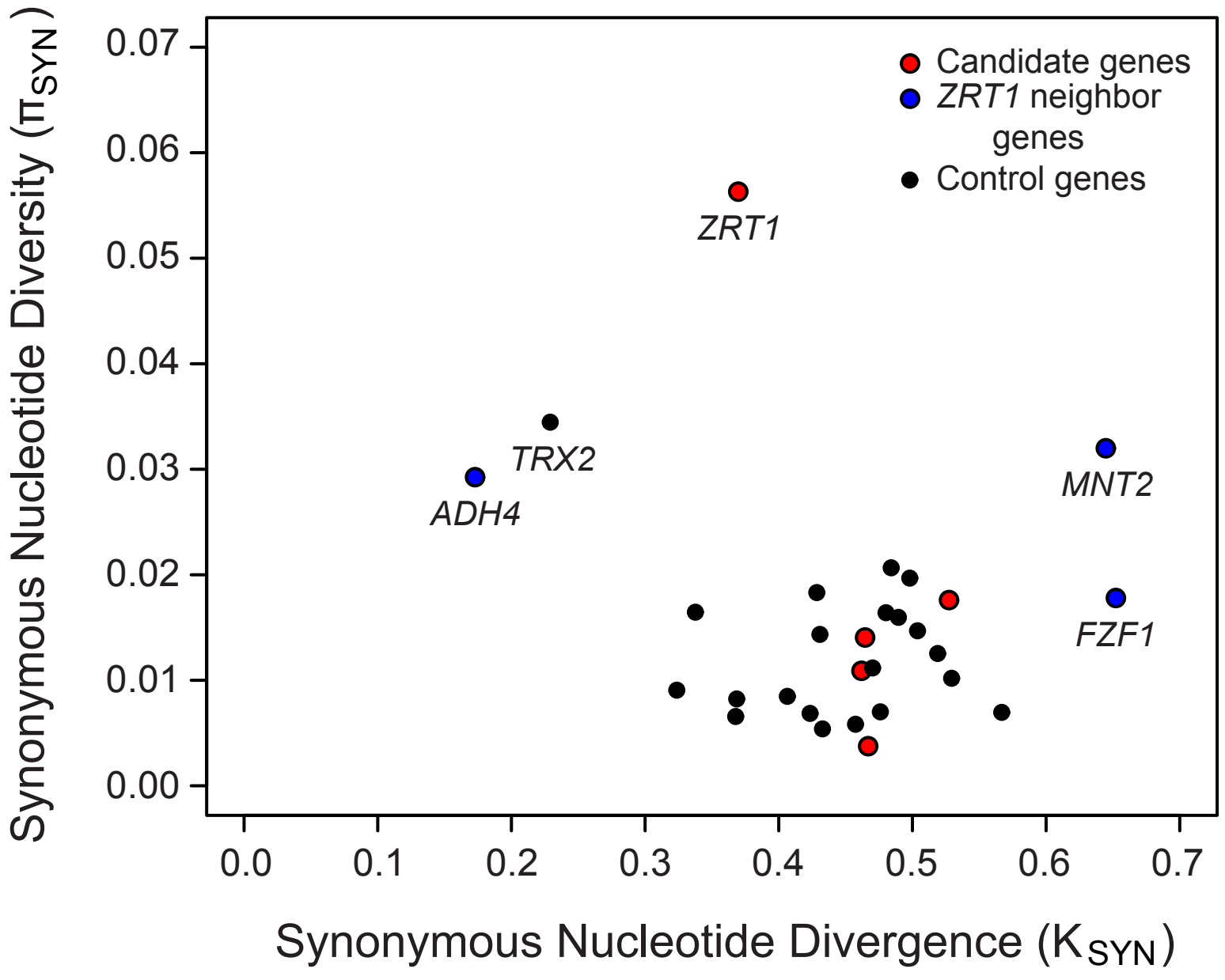

Figure 2

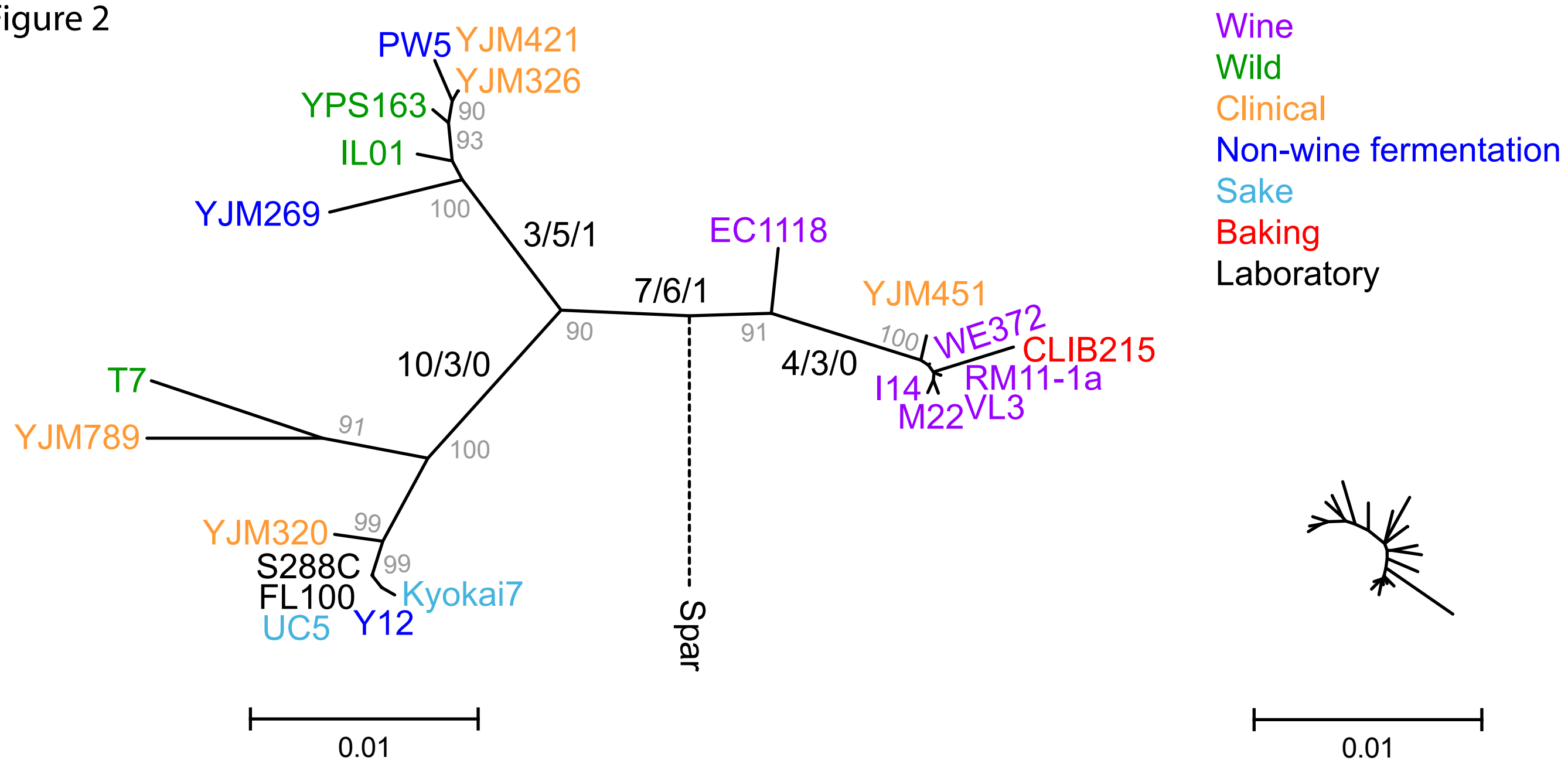

Figure 3

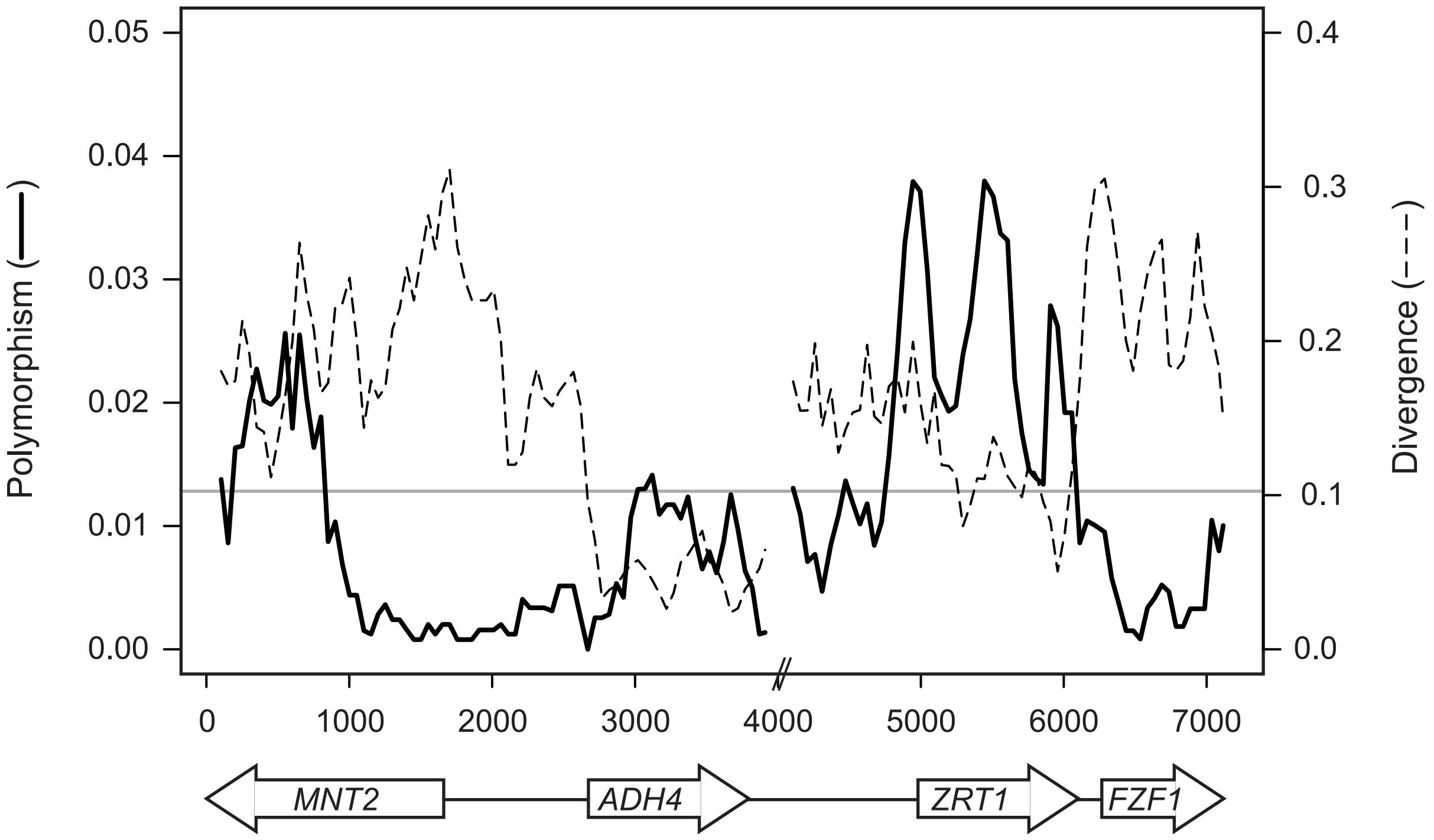

Figure 4

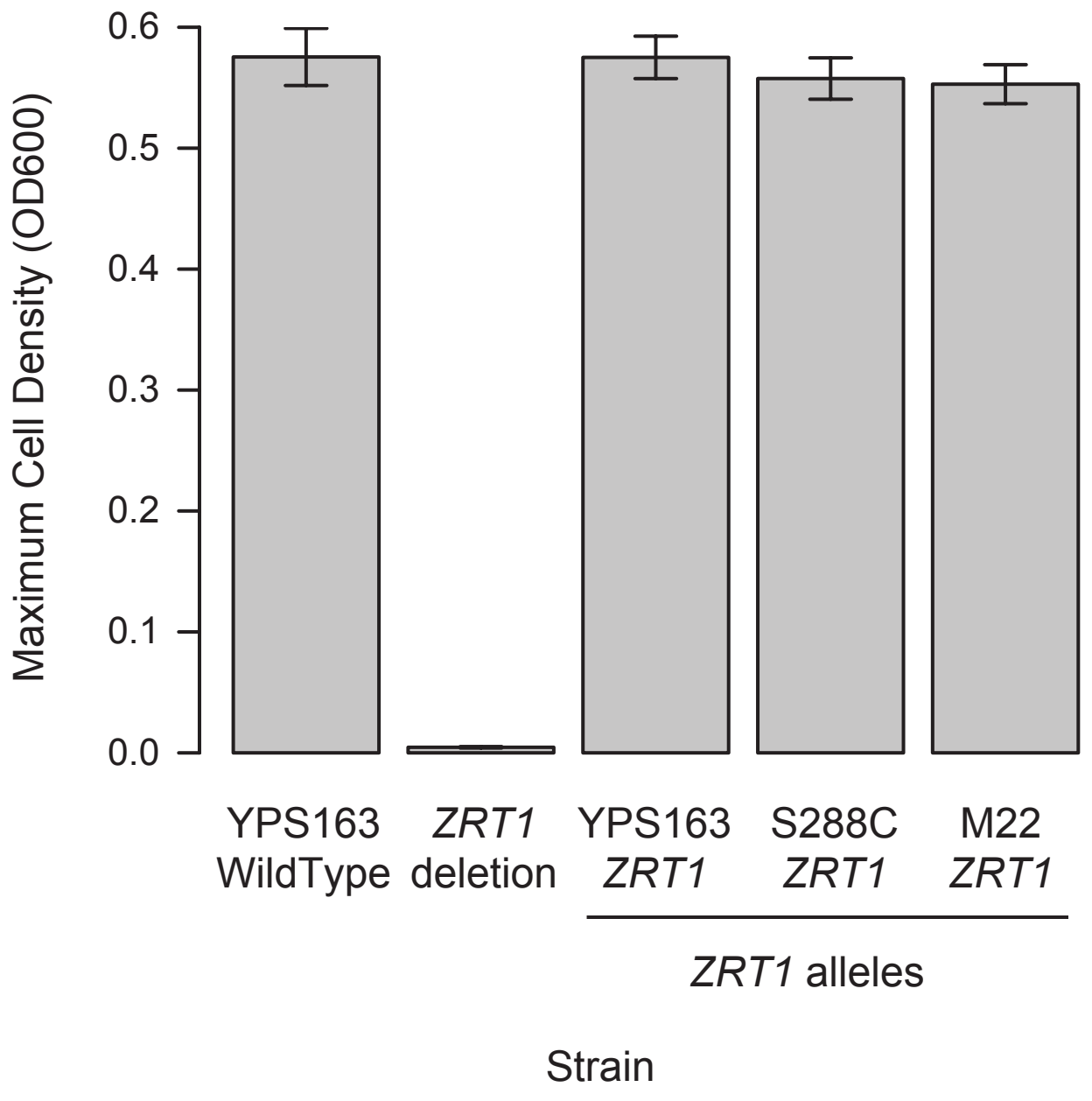

Figure 5

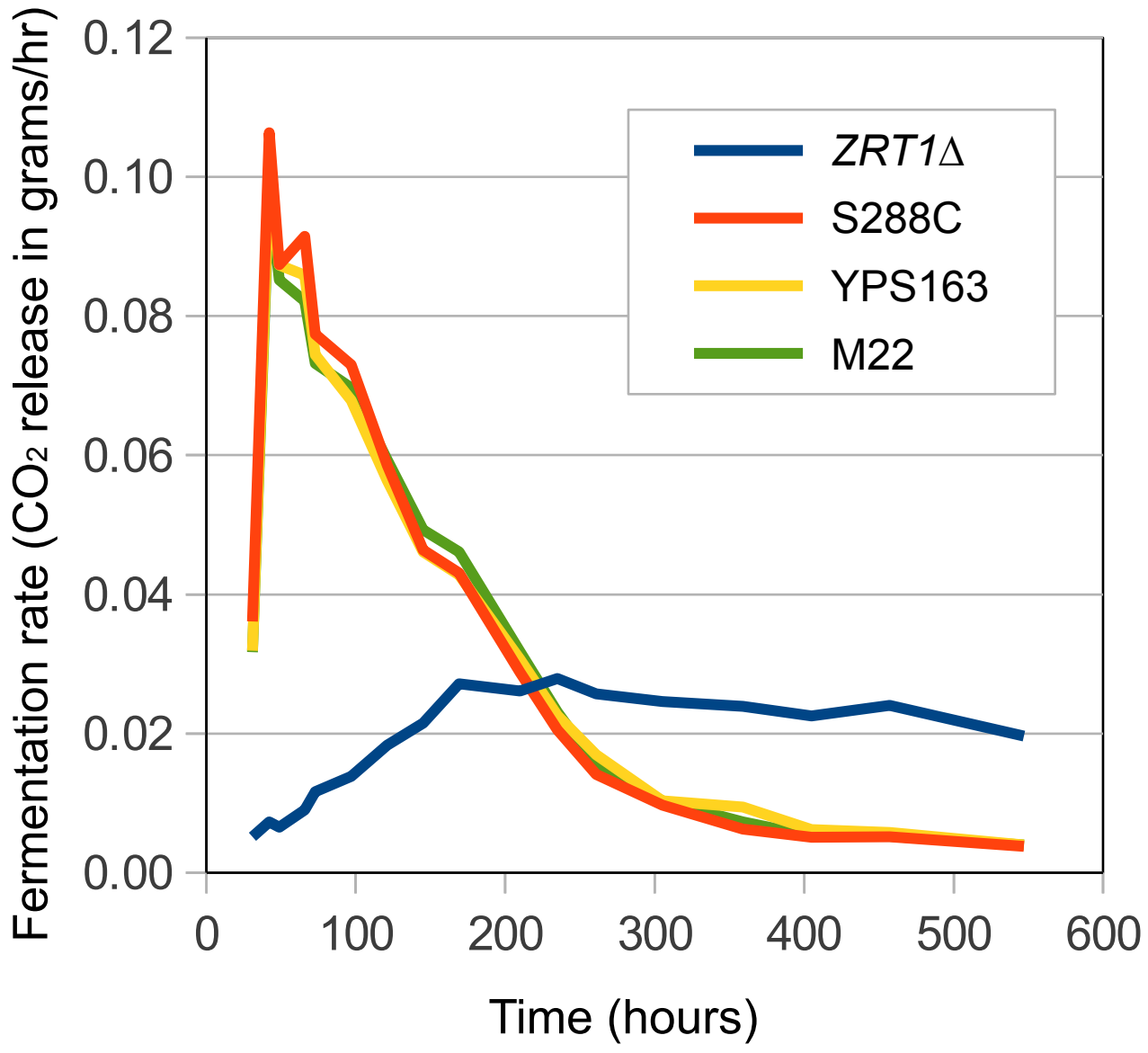

# Figure 6

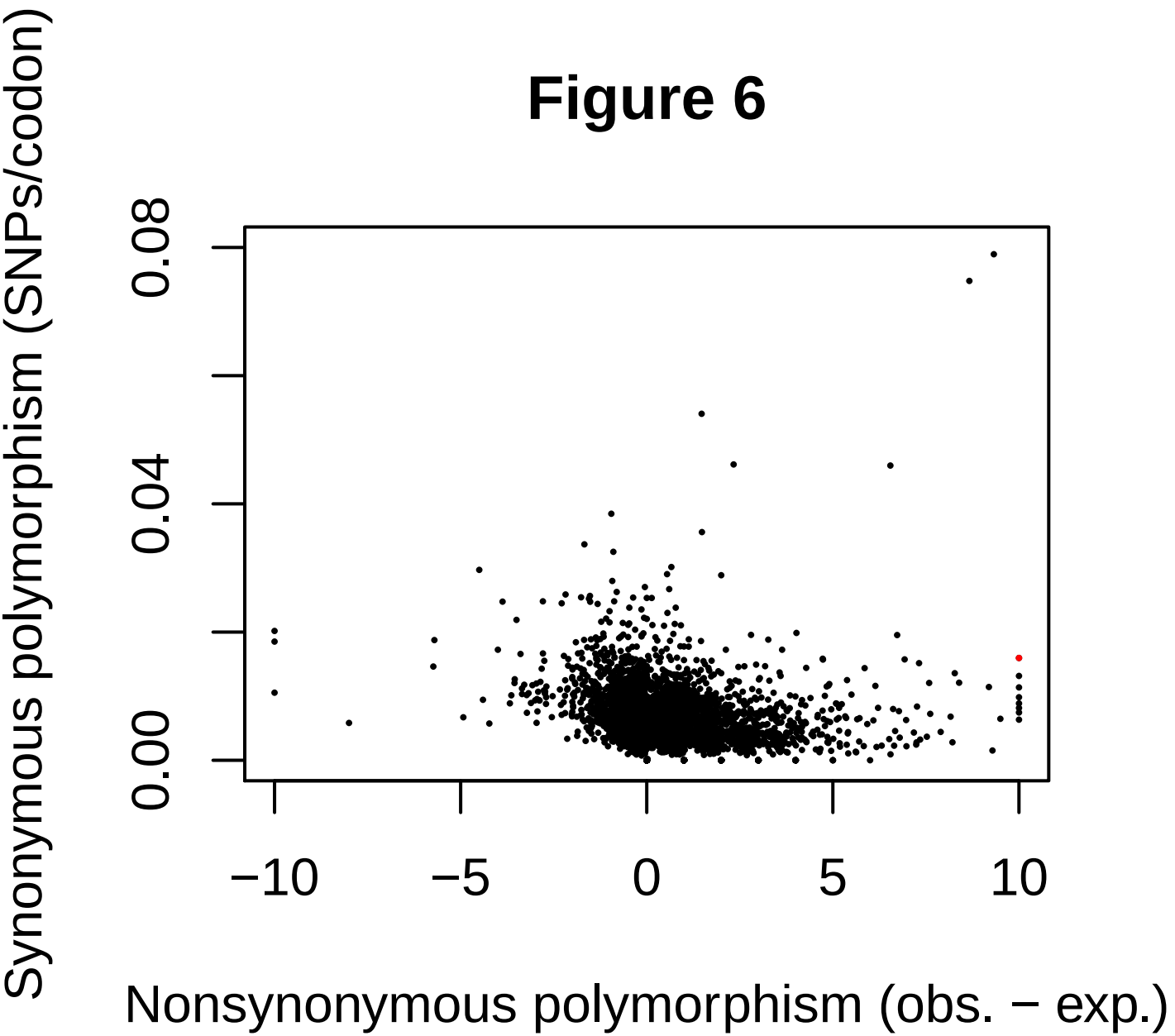

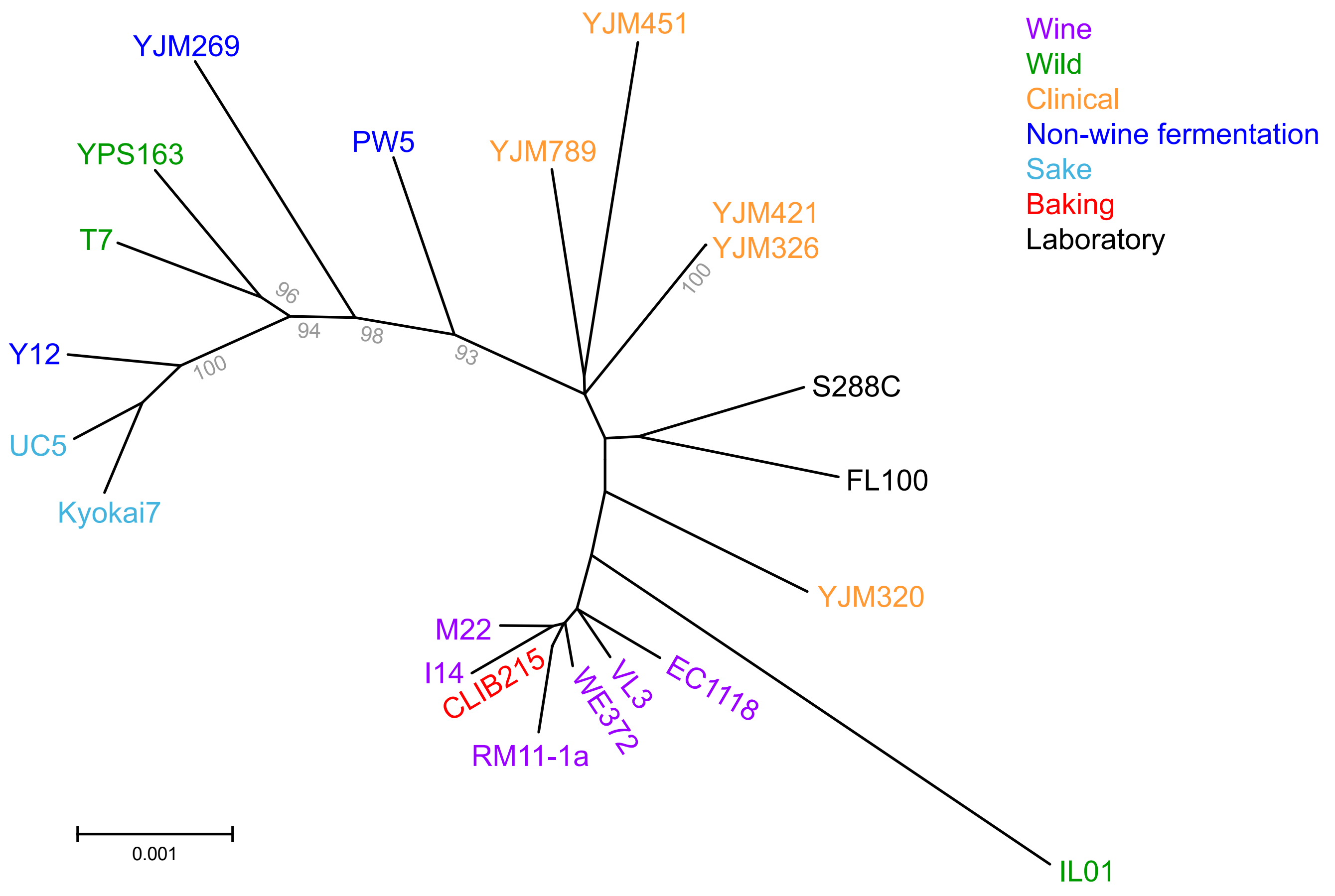

Supplementary Figure 1. Neighbor-joining tree of 21 concatenated control genes. An unrooted neighbor-joining tree of the concatenated 21 control genes along with bootstrap values greater than 90% (in gray). Only strains used in the *ZRT1* tree are shown. *S. cerevisiae* strains are color coded by strain class (see legend).

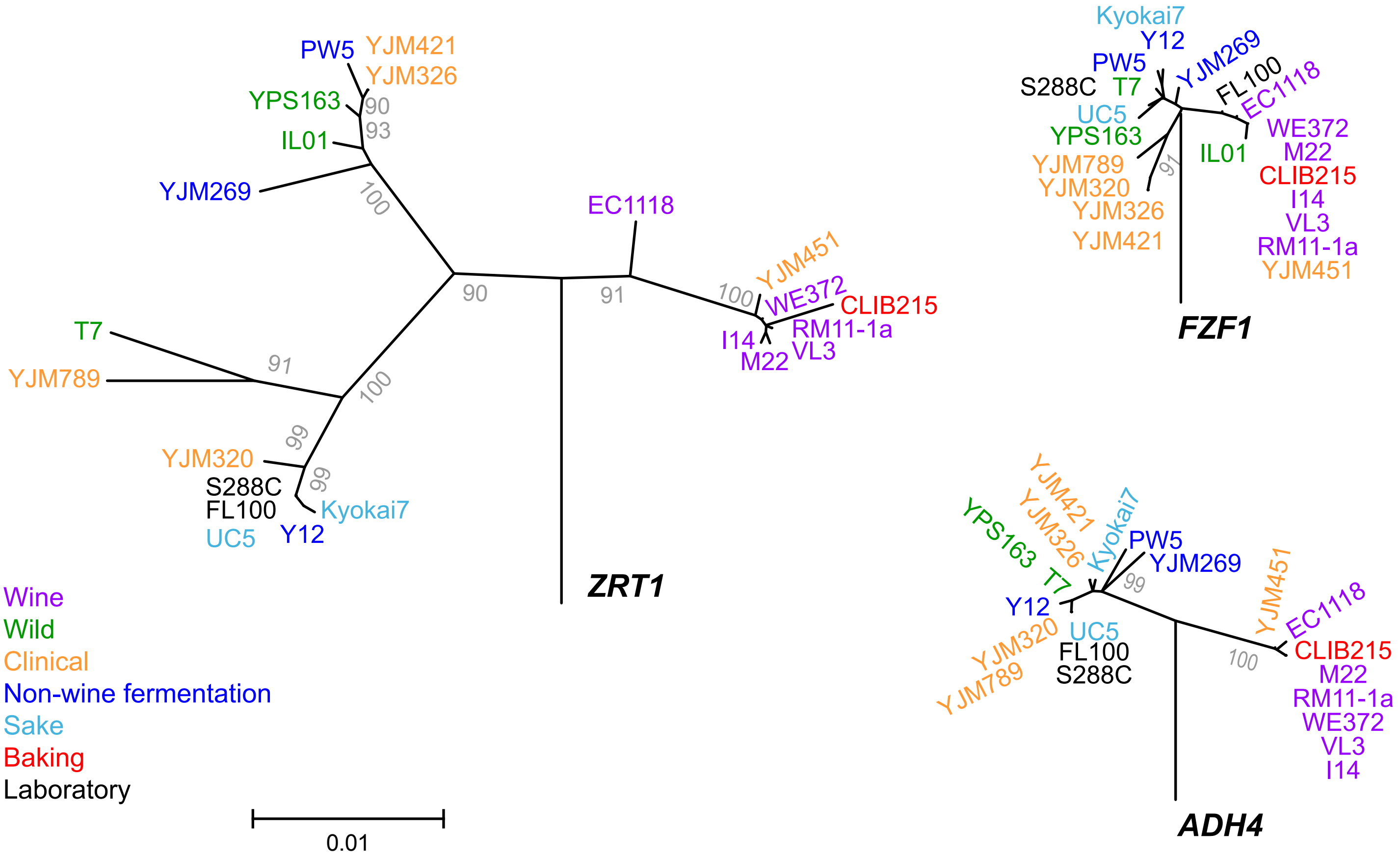

Supplementary Figure 2. Neighbor-joining tree of *ZRT1* and adjacent genes *ADH4* and *FZF1*. Neighbor-joining trees of *ZRT1* and the two adjacent genes *ADH4* and *FZF1* with bootstrap values greater than 90% (in gray) and rooted to *S. paradoxus*. *S. cerevisiae* strains are color coded by class (see legend). The strains represented in each tree are identical except strain IL01 is missing for the gene *ADH4*.

Table S1. Genome sequences used in this s

| Strain | Class |
| --- | --- |
| AWRI1631 | Vineyard/Wine |
| AWRI796 | Vineyard/Wine |
| EC1118* | Vineyard/Wine |
| Jay291 | Fermentation |
| Kyokai7 | Sake |
| LalvinAQ23 | Vineyard/Wine |
| RM11-1a | Vineyard/Wine |
| Sigma1278b | Laboratory |
| Vin13 | Vineyard/Wine |
| VL3 | Vineyard/Wine |
| YJM789 | Clinical |
| CBS7960 | Fermentation |
| CLIB215 | Baking |
| CLIB324 | Baking |
| FL100 | Laboratory |
| I14 | Vineyard/Wine |
| IL-01 | Nature |
| M22 | Vineyard/Wine |
| PW5 | Fermentation |
| T7 | Nature |
| T73 | Vineyard/Wine |
| UC5 (UCD612) | Sake |
| WE372 | Vineyard/Wine |
| Y10 (NRRL y7567) | Nature |
| Y12 (NRRL y12633) | Palm wine |
| Y9 (NRRL y5997) | Rice fermentation |
| YJM269 | Vineyard/Wine |
| YJM280 | Clinical |
| YJM320 | Clinical |
| YJM326 | Clinical |
| YJM421 | Clinical |
| YJM451 | Clinical |
| YJM653 | Clinical |
| YPS1009 | Nature |
| YPS163 | Nature |
| S288C | Laboratory |

SGD is the Saccharomyces Genome Datab
(http://www.cbi.nlm.nih.gov/), WashU is the
*based on a pseudohaploid genome

study.

| Description |
|---|
| Industrial wine strain |
| Industrial wine strain, South Africa |
| Commercial wine strain, isolated in Champagne, France |
| Biofuel sugar cane fermentation, Brazil |
| Industrial sake strain |
| Industrial wine strain |
| California, USA |
| Laboratory strain |
| Industrial wine |
| Industrial wine |
| Isolated from lung of AIDS patient with pneumonia, Kansas City, Kansas, USA |
| Factory producing ethanol from cane-sugar syrup, Sao Paulo, Brazil |
| Baker's yeast, New Zealand, 1994 |
| Baker's yeast, Saigon, Vietnam, 1996 |
| Laboratory strain |
| Vineyard soil sample, Petrina, Italy, 2002 |
| Soil sample, Cahokia, Illinois, USA, 2003 |
| Vineyard, Italy |
| Raphia Palm tree, Aba, Abia state, Nigeria, 2002 |
| Oak tree exudate, Babler State Park, Missouri, USA, 2003 |
| Monastrel grape fermentation, Alicante, Spain, 1987 |
| Isolated from Sene sake, Kurashi, Japan, pre-1974 |
| From wine in Cape Town, South Africa |
| From a coconut in the Phillipines, pre-1973 |
| From palm wine, Ivory Coast, Africa, pre-1981 |
| From Ragi (African or finger millet), Java, Indonesia, pre-1962 |
| From Blauer Portugieser grapes, 1954 |
| Peritoneal fluid, USA, pre-1994 |
| Blood sample, USA, pre-1994 |
| United States, pre-1994 |
| Ascites fluid, USA, pre-1994 |
| Europe, pre-1994 |
| Isolated from Broncho-alveolar lavage |
| Oak exudate, New Jersey, USA, 2000 |
| Oak exudate, Pennsylvania, USA, 1999 |
| Laboratory strain originally isolated from a rotting fig, California, USA, 1937 |

ase (www.yeastgenome.org), Broad is the Broad Institute (www.broad.mit.edu), N
Washington University Genome Institute (genome.wustl.edu and www.genetics.wu

| Source | Reference |
| --- | --- |
| SGD | Borneman et al. 2008 |
| SGD | Borneman et al. 2011 |
| NCBI and SGD | Novo et al. 2009 |
| SGD | Argueso et al. 2009 |
| SGD | Akao et al. 2011 |
| SGD | Borneman et al. 2011 |
| Broad | |
| SGD | Dowell et al. 2010 |
| SGD | Borneman et al. 2011 |
| SGD | Borneman et al. 2011 |
| SGD | Wei et al. 2007 |
| WashU | |
| WashU | |
| WashU | |
| WashU | |
| WashU | |
| WashU | |
| WashU | Doniger et al. 2008 |
| WashU | |
| WashU | |
| WashU | |
| WashU | |
| WashU | |
| WashU | |
| WashU | |
| WashU | |
| WashU | |
| WashU | |
| WashU | |
| WashU | |
| WashU | |
| WashU | |
| WashU | |
| WashU | |
| WashU | Doniger et al. 2008 |
| SGD | |

CBI is the National Center of Biotechnolormation
ustl.edu/jflab/data4.html).

Table S2. Polymorphism and divergence data.

| Gene set | Systematic name | Gene name | Strain sample size | S288C length | Total sites analyzed | Synonmyous sites analyzed | Ps | Pn |
|---|---|---|---|---|---|---|---|---|
| 5 gene set | YGL255W | ZRT1 | 22 | 1131 | 1113 | 267.6 | 45 | 55 |
| 5 gene set | YPR194C | OPT2 | 22 | 2634 | 2436 | 554.7 | 20 | 35 |
| 5 gene set | YDL153C | SAS10 | 24 | 1833 | 1680 | 342.5 | 7 | 12 |
| 5 gene set | YBL017C | PEP1 | 18 | 4740 | 4293 | 939.3 | 45 | 48 |
| 5 gene set | YOL081W | IRA2 | 21 | 9240 | 9117 | 2025.1 | 108 | 69 |
| ZRT1 region | YGL254W | FZF1 | 29 | 900 | 897 | 191.9 | 13 | 13 |
| ZRT1 region | YGL256W | ADH4 | 28 | 1149 | 1146 | 264.6 | 24 | 9 |
| ZRT1 region | YGL257C | MNT2 | 28 | 1677 | 1662 | 358.1 | 30 | 24 |
| 21 control | YBL068W | PRS4 | 32 | 981 | 978 | 224.4 | 12 | 1 |
| 21 control | YBR290W | BSD2 | 29 | 966 | 960 | 208.5 | 8 | 5 |
| 21 control | YDR089W | YDR089W | 23 | 2610 | 2595 | 569.7 | 26 | 21 |
| 21 control | YDR194C | ENT5 | 24 | 1995 | 1974 | 423.7 | 27 | 11 |
| 21 control | YFR043C | IRC6 | 36 | 714 | 711 | 145.3 | 4 | 10 |
| 21 control | YGR040W | KSS1 | 30 | 1107 | 1104 | 237.6 | 8 | 4 |
| 21 control | YGR209C | TRX2 | 34 | 315 | 312 | 72.4 | 8 | 3 |
| 21 control | YGR230W | BNS1 | 32 | 414 | 411 | 90.6 | 5 | 8 |
| 21 control | YGR244C | LSC2 | 26 | 1284 | 1281 | 284.6 | 12 | 6 |
| 21 control | YIL139C | REV7 | 33 | 738 | 732 | 150.4 | 7 | 4 |
| 21 control | YKL196C | YKT6 | 31 | 603 | 600 | 129.7 | 7 | 0 |
| 21 control | YKL214C | YRA2 | 32 | 612 | 600 | 130.5 | 6 | 7 |
| 21 control | YKR080W | MTD1 | 33 | 963 | 960 | 215.0 | 8 | 4 |
| 21 control | YMR065W | KAR5 | 27 | 1515 | 1503 | 319.5 | 15 | 6 |
| 21 control | YMR096W | SNZ1 | 28 | 894 | 888 | 198.8 | 5 | 6 |
| 21 control | YMR180C | CTL1 | 29 | 963 | 948 | 206.4 | 12 | 7 |
| 21 control | YMR206W | YMR206W | 29 | 942 | 936 | 212.0 | 14 | 12 |
| 21 control | YNL317W | PFS2 | 29 | 1398 | 1392 | 301.9 | 18 | 5 |
| 21 control | YNR028W | CPR8 | 33 | 927 | 918 | 204.5 | 13 | 11 |
| 21 control | YNR059W | MNT4 | 30 | 1743 | 1731 | 378.1 | 22 | 15 |
| 21 control | YPR186C | PZF1 | 28 | 1290 | 1281 | 266.2 | 23 | 12 |

| Ds | Dn | Synonymous nucleotide diversity (π) | Synonymous nucleotide divergence (K) | Tajima's D (syn sites) | Pi(n)/Pi(s ) | K(n)/K(s) | MK P-value | Neutrality Index |
|---|---|---|---|---|---|---|---|---|
| 62 | 12 | 0.056 | 0.370 | 0.72 | 0.420 | 0.114 | 0.000 | 6.31 |
| 187 | 18 | 0.011 | 0.462 | 0.59 | 0.454 | 0.029 | 0.000 | 18.18 |
| 120 | 41 | 0.004 | 0.467 | -0.98 | 0.457 | 0.076 | 0.002 | 5.02 |
| 320 | 168 | 0.014 | 0.465 | 0.04 | 0.381 | 0.129 | 0.002 | 2.03 |
| 735 | 131 | 0.018 | 0.527 | 0.76 | 0.125 | 0.041 | 0.000 | 3.58 |
| 80 | 78 | 0.018 | 0.652 | 0.11 | 0.178 | 0.193 | 1.000 | 1.03 |
| 34 | 7 | 0.029 | 0.173 | 0.95 | 0.112 | 0.065 | 0.396 | 1.82 |
| 146 | 122 | 0.032 | 0.645 | 1.96 | 0.176 | 0.168 | 1.000 | 0.96 |
| 64 | 1 | 0.008 | 0.369 | -1.22 | 0.019 | 0.004 | 0.307 | 5.33 |
| 64 | 18 | 0.008 | 0.407 | -0.41 | 0.064 | 0.062 | 0.293 | 2.22 |
| 204 | 78 | 0.013 | 0.519 | 0.05 | 0.126 | 0.079 | 0.025 | 2.11 |
| 147 | 45 | 0.015 | 0.504 | -0.52 | 0.094 | 0.061 | 0.535 | 1.33 |
| 55 | 40 | 0.010 | 0.529 | 1.31 | 0.359 | 0.154 | 0.048 | 3.44 |
| 93 | 1 | 0.007 | 0.567 | -2.63 | 0.044 | 0.004 | 0.000 | 46.50 |
| 12 | 4 | 0.034 | 0.229 | 1.31 | 0.039 | 0.077 | 1.000 | 1.13 |
| 32 | 41 | 0.021 | 0.484 | 1.36 | 0.126 | 0.308 | 0.771 | 1.25 |
| 90 | 5 | 0.005 | 0.433 | -1.72 | 0.099 | 0.012 | 0.002 | 9.00 |
| 51 | 20 | 0.006 | 0.457 | -1.42 | 0.268 | 0.083 | 0.723 | 1.46 |
| 43 | 0 | 0.011 | 0.470 | -0.51 | 0.000 | 0.000 | NA | NA |
| 32 | 14 | 0.009 | 0.324 | -0.57 | 0.358 | 0.102 | 0.189 | 2.67 |
| 61 | 6 | 0.007 | 0.368 | -0.84 | 0.061 | 0.023 | 0.040 | 5.08 |
| 111 | 77 | 0.007 | 0.476 | -1.46 | 0.131 | 0.151 | 0.350 | 0.58 |
| 62 | 16 | 0.007 | 0.423 | 0.18 | 0.220 | 0.062 | 0.024 | 4.65 |
| 71 | 92 | 0.016 | 0.489 | 0.26 | 0.151 | 0.288 | 0.144 | 0.45 |
| 55 | 34 | 0.016 | 0.338 | -0.07 | 0.168 | 0.152 | 0.501 | 1.39 |
| 103 | 9 | 0.016 | 0.480 | 0.28 | 0.023 | 0.018 | 0.127 | 3.18 |
| 66 | 75 | 0.014 | 0.431 | -0.27 | 0.159 | 0.284 | 0.517 | 0.74 |
| 116 | 92 | 0.018 | 0.428 | 1.15 | 0.140 | 0.173 | 0.722 | 0.86 |
| 91 | 43 | 0.020 | 0.498 | -0.41 | 0.115 | 0.097 | 0.841 | 1.10 |

Table S3. Maximum likelihood HKA test results.

| Gene | k[a] | P[b] |
|---|---|---|
| ZRT1 | 4.95 | 0.0000 |
| PEP1 | 0.96 | 0.4486 |
| SAS10 | 0.35 | 0.0610 |
| IRA2 | 1.03 | 0.6891 |
| OPT2 | 0.75 | 0.7163 |
| ADH4 | 4.44 | 0.0007 |
| FZF1 | 0.97 | 0.6046 |
| MNT2 | 1.30 | 0.5042 |

Each gene was tested in comparison to 21 genes in the neutral gene set.

[a] Maximum likelihood estimate of the selection parameter (k > 1 indicates an excess of polymorphisı

[b] The P-value is calculated from the chi-square distribution based on twice the difference in log-likeli

m relative to divergence)
hood between the neutral vs selection model for each gene.